\documentclass[cameraready]{Interspeech}

\title{SGAD: A State-Guided Adaptive Decision Framework for Robust EEG-Based Auditory Attention Switch Decoding}

\author[affiliation={1}]{Yuting}{Ding}
\author[affiliation={1}]{Xuefei}{Wang}
\author[affiliation={1}]{Ximin}{Chen}
\author[affiliation={2}]{Chunlin}{Li}
\author[affiliation={1}, correspondingauthor]{Fei}{Chen}

\address{
    $^1$ Department of Electronic and Electrical Engineering, Southern University of Science and Technology, Shenzhen 518055, China\\
    $^2$ School of Biomedical Engineering, Capital Medical University, Beijing 100069, China
}

\email{fchen@sustech.edu.cn}

\keywords{EEG, auditory attention switch decoding, adaptive decision-making, cross-condition evaluation}

\usepackage{comment}
\usepackage{booktabs}
\usepackage{multirow}
\usepackage{tabularx}
\newcolumntype{C}{>{\centering\arraybackslash}X} 

\begin{document}
\maketitle
\begin{abstract}
Achieving robust EEG-based auditory attention switch decoding (AASD) is crucial for intelligent hearing aids. However, its application is limited as EEG non-stationarity complicates sequential decision-making, and insufficient control of potential confounding factors may overestimate performance. Therefore, we propose a state-guided adaptive decision (SGAD) framework that infers attention transition states via causal state detection and dynamically modulates temporal smoothing through state-guided adaptive gating. We further introduce six hierarchical evaluation protocols to assess generalization across audio, speaker, and subject dimensions. Experimental results show that SGAD improves decoding accuracy and stability while maintaining low response latency across evaluation scenarios. Performance variations across protocols further suggest data partition-related biases. Together, these findings advance robust AASD for neuro-steered hearing applications.
\end{abstract}

\section{Introduction}
Individuals with normal hearing can selectively attend to a target speaker in noisy environments, a phenomenon known as the “cocktail party effect” \cite{cherry1953some,ding2012neural}. However, people with hearing loss often struggle with this ability, posing a core challenge for hearing-assistive devices \cite{haykin2005cocktail}. With advances in neuroscience, auditory attention decoding (AAD) has emerged as a promising approach for inferring attended speech from the electroencephalography (EEG) signal, thereby enabling neuro-steered hearing systems in multi-talker scenarios \cite{aad4}. Despite this potential, early AAD studies primarily focused on static-attention paradigms \cite{puffay2023relating}, whereas attention in real-world listening environments shifts dynamically with contextual demands and cognitive load \cite{bronkhorst2015cocktail}, limiting the applicability of such approaches in practical settings. Consequently, recent research has shifted toward auditory attention switch decoding (AASD) to better capture dynamic attentional transitions \cite{teoh2019eeg,ivucic2025selective,VANDERYCK2026109539}.

In AASD settings, reliable decoding relies on continuous and temporally stable tracking of auditory attention as it shifts between competing speakers \cite{geirnaert2021electroencephalography,schiavon2026impact}. While recent studies have improved window-wise discrimination by refining EEG encoders for feature extraction, they typically generate independent predictions from short temporal segments without explicitly modeling temporal dependencies across consecutive windows \cite{fan2025seeing,fan2025listennet,ding2025eeg}. As a result, window-wise decoding can introduce substantial variability in the outputs, particularly during attentional transitions due to the inherent non-stationarity of EEG \cite{brain,rotaru2024we}. To mitigate these fluctuations, simple sequential decision strategies, such as fixed temporal smoothing, are frequently adopted as a baseline post-processing step after feature extraction to stabilize predictions. However, since this time-invariant smoothing is applied uniformly across both stable and transitional states, it introduces an inherent trade-off between decoding stability and tracking speed \cite{hjortkjaer2025real}. Therefore, more effective sequential decision strategies capable of state-dependent temporal modulation are required to ensure temporal consistency, together with comprehensive evaluation across decoding accuracy, switch detection capability, and response latency.

In addition to sequential decision-making, reliable evaluation of model generalization across diverse acoustic environments is essential for practical AASD deployment \cite{puffay2023relating}. However, such assessment remains challenging due to non-attentional cues introduced by different data partitioning, which data-driven models are prone to overfit and exploit as confounding factors, thereby inflating decoding performance \cite{rotaru2024we,qiu2025enhancing,ding2025developing}. Although cross-trial and cross-subject evaluations are widely adopted to mitigate partition-related biases, they often overlook potential confounding effects of audio content and speaker identity, limiting reliable assessment of model generalization under realistic conditions \cite{cai2024robust,zhang2024swim,zhang2025multi}. Consequently, relying on a single partitioning protocol is insufficient to rigorously evaluate model robustness, motivating evaluation under multiple hierarchical generalization constraints.

To bridge these gaps, we propose a state-guided adaptive decision (SGAD) framework together with structured evaluation protocols, which integrates robust sequential decision-making with rigorous generalization assessment for reliable AASD. The main contributions are summarized as follows:

\begin{itemize}
\item The proposed SGAD framework incorporates a causal state detector to estimate the attention-switch probability and performs state-guided temporal smoothing via an adaptive gating mechanism. By leveraging state-dependent temporal modulation, it suppresses prediction fluctuations during steady states while reducing response latency during attentional transitions. As a decision-level module, it is encoder-agnostic and compatible with diverse decoding backbones.

\item Six hierarchical evaluation protocols are established by combining three partition factors (trial, audio content, and speaker identity) with two subject conditions (subject-dependent and subject-independent), enabling rigorous analysis of model generalization under diverse constraints.

\item Experimental results demonstrate that the proposed SGAD improves decoding accuracy and stability while effectively controlling response latency, and the structured evaluation further highlights the importance of comprehensive evaluation schemes for reliable AASD assessment.
\end{itemize}

\section{Methods}

\begin{figure*}[t]
  \centering
  \includegraphics[width=0.9\linewidth]{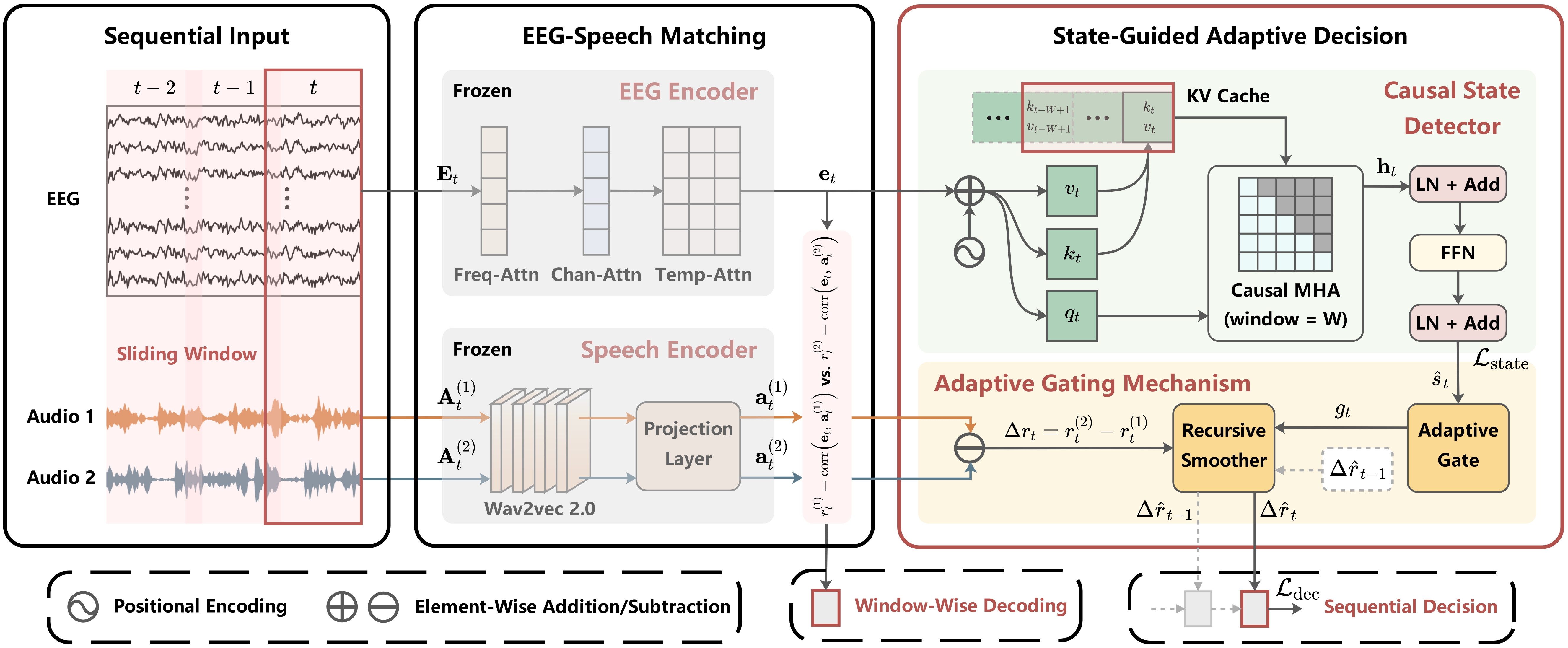}
  \caption{Architecture of the proposed SGAD-based sequential decision framework with three stages: (i) sequential input partitioning via sliding windows, (ii) EEG-speech matching via frozen encoders, and (iii) state-guided adaptive decision making.}
  \label{fig:SGAD}
\end{figure*}

\subsection{EEG-speech matching}
To provide attention evidence for sequential decision-making, we first formulate AASD as an EEG-speech matching problem. For each time $t$, the multi-band EEG segment $\mathbf{E}_t \in \mathbb{R}^{N \times C \times T}$, where $N$, $C$, and $T$ denote the number of sub-bands, channels, and time steps, respectively, is mapped by an EEG encoder $f_{\mathrm{EEG}}(\cdot)$ to $\mathbf{e}_t \in \mathbb{R}^{d}$. Similarly, the two competing speech segments $\mathbf{A}_t^{(1)}, \mathbf{A}t^{(2)} \in \mathbb{R}^{d \times T}$,
are transformed by a speech encoder $f_{\mathrm{speech}}(\cdot)$ into $\mathbf{a}_t^{(1)}, \mathbf{a}_t^{(2)} \in \mathbb{R}^{d}$, where $d=64$. We implement $f_{\mathrm{EEG}}(\cdot)$ using CNNT \cite{bollens2024contrastive} and FCTNet \cite{ding2025eeg} architectures, while $f_{\mathrm{speech}}(\cdot)$ utilizes a pretrained wav2vec 2.0 model \cite{baevski2020wav2vec} with a $1\times1$ convolutional projection layer. Alignment is quantified by the Pearson correlation $r_t^{(i)} = \mathrm{corr}(\mathbf{e}_t, \mathbf{a}_t^{(i)})$. The encoders are pretrained using a cross-entropy loss $\mathcal{L}_{\mathrm{match}} = \mathrm{CE}([ r_t^{(1)}, r_t^{(2)} ], y_t)$, where $y_t \in \{1,2\}$ denotes the ground-truth attended speaker.

\subsection{Proposed SGAD framework}
Based on the raw correlation scores $r_t^{(i)}$, a straightforward approach is to perform \textit{window-wise decoding} (WD) as $\hat{y}_t=\arg\max_i r_t^{(i)}$. However, operating independently at each window makes WD highly sensitive to EEG non-stationarity and unsuitable for stable AASD. To ensure temporal consistency, sequential decision strategies such as \textit{persistence-based decision} (PBD) and \textit{exponential moving average (EMA)-based decision} (EBD) apply fixed temporal smoothing across consecutive windows. Specifically, PBD confirms an attention switch only after $K$ consecutive identical predictions, whereas EBD smooths the decision margin $\Delta r_t = r_t^{(2)} - r_t^{(1)}$ using a fixed EMA,
$\Delta \hat{r}_t = \alpha \Delta \hat{r}_{t-1} + (1-\alpha)\Delta r_t$. Although PBD and EBD suppress short-term fluctuations, they rely on static temporal constraints that cannot differentiate between stable and transition states, thereby slowing adaptation to attention switches. Therefore, as shown in Fig.~\ref{fig:SGAD}, we propose a state-guided adaptive decision (SGAD) framework, which adaptively modulates temporal smoothing based on the estimated transition state.

\subsubsection{Causal state detector (CSD)}
Given that attentional transitions evolve gradually over time, relying on single-window evidence is inherently unreliable for switch detection \cite{puffay2023relating,brain}. The CSD therefore models causal temporal dependencies across multiple preceding windows to capture the evolving attention state rather than isolated fluctuations. At each time step $t$, the EEG embedding $\mathbf{e}_t$ is combined with positional encoding $\mathbf{p}_t$ as $\tilde{\mathbf{e}}_t=\mathbf{e}_t+\mathbf{p}_t \in \mathbb{R}^{d}$, and then projected into query, key, and value representations:
\begin{equation}
\mathbf{q}_t = \tilde{\mathbf{e}}_t \mathbf{W}_Q, \quad
\mathbf{k}_t = \tilde{\mathbf{e}}_t \mathbf{W}_K, \quad
\mathbf{v}_t = \tilde{\mathbf{e}}_t \mathbf{W}_V,
\end{equation}
where $\mathbf{W}_Q, \mathbf{W}_K, \mathbf{W}_V \in \mathbb{R}^{d \times d_k}$ are learnable projection matrices and $d_k=16$. To ensure causal inference, a key-value (KV) cache maintains historical keys and values within a sliding window of length $W$, enabling causal multi-head attention (MHA) \cite{vaswani2017attention}. The contextual feature is then computed as:
\begin{equation}
\mathbf{h}_t =
\mathrm{Softmax}
\left(
\frac{\mathbf{q}_t \mathbf{k}_{t-W+1:t}^\top}{\sqrt{d_k}}
\right)
\mathbf{v}_{t-W+1:t},
\end{equation}
where $\mathbf{k}_{t-W+1:t}$ and $\mathbf{v}_{t-W+1:t}$ denote the cached keys and values. The attention output is further processed as:
\begin{equation}
\mathbf{u}_t = \mathrm{LN}(\mathbf{h}_t + \tilde{\mathbf{e}}_t), \quad
\mathbf{z}_t = \mathrm{LN}(\mathrm{FFN}(\mathbf{u}_t) + \mathbf{u}_t),
\end{equation}
where $\mathrm{LN}(\cdot)$ denotes layer normalization and $\mathrm{FFN}(\cdot)$ is a position-wise feed-forward network. The transition probability is computed as $\hat{s}_t=\sigma(\mathbf{z}_t)$, where $\hat{s}_t\in[0,1]$ and $\sigma(\cdot)$ denotes a sigmoid function.

\subsubsection{Adaptive gating mechanism (AGM)}
Guided by the transition probability $\hat{s}_t$, the AGM dynamically regulates decision updates. The adaptive gate maps $\hat{s}_t$ to a smoothing factor $g_t$ via a monotonic decreasing function:
\begin{equation}
g_t = g_{\min} + (g_{\max}-g_{\min}) \cdot \sigma(b-a\hat{s}_t),
\end{equation}
where $a>0$ and $b$ are learnable parameters controlling the gating function. The bounds $g_{\min}=0.05$ and $g_{\max}=0.95$ constrain the smoothing factor within a practical range to balance tracking agility and noise suppression. This formulation ensures that higher $\hat{s}_t$ yields smaller $g_t$, reducing system inertia and enabling rapid adaptation during attention switches, whereas lower $\hat{s}_t$ produces larger $g_t$ to enforce stronger temporal smoothing in stable periods. Subsequently, the recursive smoother updates the decision margin $\Delta \hat{r}_t$ using the current evidence $\Delta r_t = r_t^{(2)} - r_t^{(1)}$ and the previous state $\Delta \hat{r}_{t-1}$:
\begin{equation}
\Delta \hat{r}_t
=
g_t \Delta \hat{r}_{t-1}
+
(1-g_t)\Delta r_t.
\end{equation}
By integrating this state-guided gain into the recursive loop, SGAD decouples the fixed temporal trade-off inherent in conventional EBD methods.

\subsubsection{Training objectives}
With frozen encoders, SGAD is trained via a multi-task objective:
\begin{equation}
\mathcal{L}_{\text{SGAD}}
=
\mathcal{L}_{\text{dec}}
+
\lambda_1 \mathcal{L}_{\text{state}}
+
\lambda_2 \mathcal{L}_{\text{smooth}},
\end{equation}
where $\lambda_1=0.5$ and $\lambda_2=0.05$ based on validation experiments. The primary objective for attention decoding is $\mathcal{L}_{\text{dec}}=\mathrm{BCE}(\sigma(\Delta\hat{r}_t),y_t)$, which optimizes the filtered decision margin. To provide additional explicit supervision for the CSD module in capturing dynamic attention transitions, an auxiliary state loss $\mathcal{L}_{\text{state}}=\mathrm{BCE}(\hat{s}_t,s_t)$ is introduced. Here, triangular soft labels are defined as $s_t=\max(0,1-|t-t_{\text{sw}}|/R)$, where $t_{\text{sw}}$ denotes the ground-truth switch time and $R=5$~s avoids sparse state supervision, yielding non-zero labels for about half the windows. Finally, $\mathcal{L}_{\text{smooth}}=\sum_t(g_t-g_{t-1})^2$ regularizes temporal fluctuations of the adaptive gate.

\section{Experiments}
\subsection{Dataset and preprocessing}
Experiments were conducted on the MS-AASD dataset \cite{ding2026saasdnet, ding_2025_17149387} with EEG recordings from 13 normal-hearing adults (aged 21--28 years) performing self-initiated attention switching in mixed-speech environments. EEG was recorded at 500~Hz using 64 channels. In each 60-s trial, two root-mean-square (RMS)-normalized speech segments from one of three male-female speaker pairs were mixed at 0~dB and presented binaurally. Participants attended to one speaker and voluntarily switched attention between concurrent streams, reporting attention states via two-button keypresses. Across 60 trials per participant, 13 hours of data were recorded, with a mean of 3.37 attention switches per trial.

EEG signals were re-referenced to the mastoid average, band-pass filtered (0.5--50~Hz) and decomposed into five sub-bands ($\delta$, $\theta$, $\alpha$, $\beta$, $\gamma$), then downsampled to 128~Hz and normalized per channel. Speech features were extracted from the 14-th layer of pretrained wav2vec~2.0 \cite{baevski2020wav2vec}, reduced to 64 dimensions via PCA, and resampled to 128~Hz. Continuous EEG and speech features were segmented into overlapping 1.0~s windows with a 0.2~s step, starting from the first annotated attention event in each trial. Each segment was labeled by its dominant attention state to mitigate sensitivity to keypress timing variability. A switch window was defined when the dominant label of the current window differed from that of the preceding window.

\subsection{Evaluation protocols}
As summarized in Table~\ref{tab:protocols}, six protocols were designed to assess model generalization of audio content, speaker identity, and subject information: \textit{leave-one-trial-out} (LOTO) \cite{zhang2025multi}, \textit{leave-one-audio-out} (LOAO) \cite{zhang2025multi}, \textit{leave-one-speaker-pair-out} (LOSpO) \cite{zhang2024swim}, \textit{leave-one-subject-out} (LOSO) \cite{zhang2024swim}, \textit{leave-one-subject-and-audio-out} (LOSAO), and \textit{leave-one-subject-and-speaker-pair-out} (LOSSO). Specifically, LOTO, LOAO, and LOSpO assess subject-dependent generalization using 3-fold cross-validation. LOTO performs random trial partitioning to evaluate cross-trial robustness, LOAO withholds shared audio folds to ensure stimulus independence, and LOSpO excludes one predefined male-female speaker pair to assess speaker independence. In contrast, LOSO, LOSAO, and LOSSO evaluate subject-independent generalization under a unified 39-fold structure ($13 \times 3$ content partitions), where content partitions are defined by audio stimuli or speaker pairs. LOSO withholds one speaker pair per fold to ensure a fair training budget comparison, LOSAO jointly enforces unseen-subject and unseen-audio conditions, and LOSSO evaluates the most rigorous joint subject-speaker independence scenario. For all protocols, five trials per subject are reserved for validation.

\begin{table}[t]
\centering
\caption{Seen and unseen conditions of the six protocols}
\label{tab:protocols}
\begin{tabular}{l ccc}
\toprule
\textbf{Protocol} & \textbf{Audio} & \textbf{Speaker} & \textbf{Subject} \\
\midrule
LOTO / LOSO   & Seen   & Seen   & Seen / Unseen \\
LOAO / LOSAO  & Unseen & Seen   & Seen / Unseen \\
LOSpO / LOSSO & Unseen & Unseen & Seen / Unseen \\
\bottomrule
\end{tabular}
\end{table}

\subsection{Experimental setup and evaluation metrics}
All models were implemented in PyTorch and trained on NVIDIA V100 GPUs using AdamW (weight decay $1\times10^{-4}$). EEG and speech encoders were pretrained for 50 epochs using $\mathcal{L}_{\mathrm{match}}$ (learning rate $1\times10^{-3}$, batch size 64, pair-swap $p=0.5$), after which they were frozen and SGAD was trained on sequential trial data using $\mathcal{L}_{\mathrm{SGAD}}$ under identical optimization settings with a two-window burn-in. Early stopping with a patience of 10 epochs was applied in both stages. Decision hyperparameters were selected via validation grid search over $K \in \{2,4,\dots,10\}$, $\alpha \in \{0.5,0.6,\dots,0.9\}$, and $W \in \{5,10,\dots,25\}$, yielding $K=5$, $\alpha=0.8$, and $W=15$.

AASD performance was evaluated using three metrics: \textit{decoding accuracy} (Acc), \textit{switch-F1 score} (Sw-F1), and \textit{switch detection latency} (SDL). Acc measures segment-level decoding performance and is defined as \cite{puffay2023relating}:
\begin{equation}
\mathrm{Acc}=\frac{1}{N}\sum_{i=1}^{N}\mathbb{I}(\hat{y}_i=y_i),
\end{equation}
where $y_i$ and $\hat{y}_i$ denote the ground-truth and predicted attention labels of the $i$-th window, respectively, and $N$ is the total number of windows. To evaluate decision stability, switch detection was formulated as an event-level task following standard EEG event detection paradigms \cite{Warby2014,shah2021objective}. A true positive switch is defined as the first prediction within a 5.0~s tolerance window (the 99th percentile of validation-set detection latency) following a ground-truth switch. Sw-F1 is defined as:
\begin{equation}
\mathrm{Sw\mbox{-}F1}
=
\frac{2N_{\mathrm{match}}}{N_{\mathrm{pred}}+N_{\mathrm{gt}}},
\end{equation}
where $N_{\mathrm{match}}$, $N_{\mathrm{pred}}$, and $N_{\mathrm{gt}}$ denote the numbers of correctly matched switches, predicted switches, and ground-truth switches, respectively. Additionally, SDL measures switching responsiveness and is computed as the average delay between matched predicted and ground-truth switches \cite{hjortkjaer2025real}:
\begin{equation}
\mathrm{SDL}=\frac{1}{N_{\mathrm{match}}}\sum_{j=1}^{N_{\mathrm{match}}}(t_{\mathrm{pred},j}-t_{\mathrm{gt},j}),
\end{equation}
where $t_{\mathrm{pred},j}$ and $t_{\mathrm{gt},j}$ are the timestamps of the $j$-th matched predicted and ground-truth switches, respectively.

\begin{table*}[t]
\centering
\caption{Performance comparison across six evaluation protocols, with Acc and Sw-F1 in percentage (\%), and SDL in seconds (s)}
\label{tab:main_results}
\begin{tabular}{ll|ccc|ccc|ccc|ccc}
\toprule
\multirow{4}{*}{\textbf{Encoder}} &
\multirow{4}{*}{\textbf{Protocol}} &
\multicolumn{12}{c}{\textbf{Decision Framework}} \\
\cmidrule(lr){3-14}
& 
& \multicolumn{3}{c}{\textbf{WD}}
& \multicolumn{3}{c}{\textbf{PBD}}
& \multicolumn{3}{c}{\textbf{EBD}}
& \multicolumn{3}{c}{\textbf{SGAD}} \\
\cmidrule(lr){3-5}
\cmidrule(lr){6-8}
\cmidrule(lr){9-11}
\cmidrule(lr){12-14}
& 
& Acc & Sw-F1 & SDL
& Acc & Sw-F1 & SDL
& Acc & Sw-F1 & SDL
& Acc & Sw-F1 & SDL \\
\midrule

\multirow{6}{*}{\textbf{CNNT}}
& LOTO
& 80.2 & 27.0 & \textbf{0.77}
& 81.3 & 45.3 & 1.08
& 82.4 & 54.6 & 1.23
& \textbf{82.8} & \textbf{70.2} & 1.04 \\

& LOAO
& 77.6 & 29.1 & \textbf{0.80}
& 78.9 & 46.8 & 1.12
& 79.5 & 56.2 & 1.37
& \textbf{80.2} & \textbf{72.4} & 1.07 \\

& LOSpO
& 75.0 & 28.4 & \textbf{0.79}
& 76.5 & 49.1 & 1.15
& 75.8 & 52.4 & 1.25
& \textbf{77.3} & \textbf{65.2} & 0.98 \\

& LOSO
& 76.2 & 27.6 & \textbf{0.87}
& 77.5 & 43.8 & 1.18
& 78.3 & 51.7 & 1.33
& \textbf{78.8} & \textbf{66.2} & 1.12 \\

& LOSAO
& 74.4 & 25.9 & \textbf{1.20}
& 75.7 & 39.7 & 1.55
& 76.1 & 47.9 & 1.68
& \textbf{76.8} & \textbf{64.0} & 1.48 \\

& LOSSO
& 72.1 & 24.8 & \textbf{1.14}
& 72.9 & 38.2 & 1.48
& 73.8 & 46.5 & 1.62
& \textbf{74.5} & \textbf{59.4} & 1.39 \\

\midrule

\multirow{6}{*}{\textbf{FCTNet}}

& LOTO
& 83.2 & 34.1 & \textbf{0.70}
& 84.4 & 50.6 & 1.10
& 84.7 & 58.2 & 1.25
& \textbf{85.6} & \textbf{72.5} & 1.12 \\

& LOAO
& 80.9 & 32.5 & \textbf{0.76}
& 82.1 & 52.8 & 1.07
& 81.5 & 60.9 & 1.21
& \textbf{82.9} & \textbf{71.3} & 1.04 \\

& LOSpO
& 78.2 & 31.8 & \textbf{0.78}
& 79.4 & 46.9 & 1.09
& 79.7 & 57.8 & 1.23
& \textbf{80.5} & \textbf{69.7} & 1.05 \\

& LOSO
& 79.7 & 31.1 & \textbf{0.81}
& 81.0 & 47.6 & 1.12
& 81.4 & 55.4 & 1.26
& \textbf{82.1} & \textbf{68.9} & 1.16 \\

& LOSAO
& 77.1 & 28.3 & \textbf{0.90}
& 78.3 & 43.8 & 1.36
& 78.7 & 51.2 & 1.49
& \textbf{79.4} & \textbf{65.8} & 1.28 \\

& LOSSO
& 73.9 & 27.4 & \textbf{1.13}
& 75.1 & 42.1 & 1.48
& 76.0 & 51.6 & 1.61
& \textbf{76.7} & \textbf{62.3} & 1.42 \\

\midrule \multicolumn{2}{c|}{\textbf{Mean (All)}} 
& 77.4 & 29.0 & \textbf{0.89} 
& 78.6 & 45.6 & 1.23 
& 79.0 & 53.7 & 1.38 
& \textbf{79.8} & \textbf{67.3} & 1.18 \\

\bottomrule
\end{tabular}
\end{table*}

\begin{table}[t]
\centering
\caption{Ablation study of SGAD under the LOSSO protocol}
\label{tab:ablation}
\setlength{\tabcolsep}{4.5pt}
\begin{tabular}{l|ccc|ccc}
\toprule
\multirow{2}{*}{\textbf{Model}} 
& \multicolumn{3}{c|}{\textbf{CNNT}} 
& \multicolumn{3}{c}{\textbf{FCTNet}} \\
\cmidrule(lr){2-4}
\cmidrule(lr){5-7}
& Acc & Sw-F1 & SDL 
& Acc & Sw-F1 & SDL \\
\midrule

Full SGAD
& \textbf{74.5} & \textbf{59.4} & 1.39
& \textbf{76.7} & \textbf{62.3} & 1.42 \\

-- w/o CSD 
& 73.1 & 46.8 & \textbf{1.25}
& 75.2 & 48.5 & \textbf{1.28} \\

-- w/o AGM 
& 73.6 & 54.0 & 1.46
& 75.8 & 56.8 & 1.49 \\

-- w/o $\mathcal{L}_{\text{state}}$  
& 74.0 & 51.2 & 1.49
& 76.1 & 53.7 & 1.52 \\

-- w/o $\mathcal{L}_{\text{smooth}}$ 
& 74.2 & 57.0 & 1.31
& 75.9 & 60.4 & 1.32 \\

\bottomrule
\end{tabular}
\end{table}

\section{Results and Discussion}

\subsection{Performance and ablation analysis}
Table~\ref{tab:main_results} compares SGAD with three baseline decision frameworks across two encoders and six evaluation protocols. WD achieves the lowest SDL of 0.89~s by operating independently on each window, but its Sw-F1 remains low at 29.0\%, indicating limited robustness in tracking attention transitions. PBD and EBD improve temporal consistency, increasing the mean Acc from 77.4\% to 78.6\% and 79.0\%, respectively, while raising Sw-F1 to 45.6\% and 53.7\%. However, their fixed smoothing increases system inertia, resulting in higher SDL values of 1.23~s and 1.38~s. In contrast, SGAD achieves a better trade-off among Acc, Sw-F1, and SDL, improving the mean Acc to 79.8\% and Sw-F1 to 67.3\% while maintaining a lower SDL of 1.18~s compared with EBD. This balanced improvement is consistently observed across all experimental settings. Overall, incorporating transition-state information into the recursive update enables SGAD to enhance switch detection performance without the substantial latency penalty inherent in fixed smoothing, thereby enabling more adaptive decision-making.

Table~\ref{tab:ablation} reports the ablation results under the LOSSO protocol. Replacing the CSD with a single-window state output markedly reduces Sw-F1 (e.g., from 62.3\% to 48.5\% for FCTNet), while achieving a lower SDL of 1.28~s, highlighting the critical role of temporal context in transition estimation. Removing the AGM and substituting the learnable gate with a fixed mapping based on $(1-\hat{s}_t)$ also degrades performance, underscoring the importance of learnable gating for adaptive smoothing. Furthermore, eliminating the auxiliary loss $\mathcal{L}_{\text{state}}$ reduces performance due to the lack of explicit transition supervision, while removing $\mathcal{L}_{\text{smooth}}$ similarly leads to performance degradation by weakening temporal regularization. Collectively, these findings highlight the complementary roles of temporal context, adaptive gating, and auxiliary supervision, confirming the importance of explicit transition-state modeling for robust sequential decision-making.

\subsection{Comparison of evaluation protocols}
Performance variations across the six protocols demonstrate the impact of generalization constraints on decoding performance. In both subject-dependent and subject-independent settings, accuracy declines as additional constraints on audio content and speaker identity are imposed. For example, under subject-dependent evaluation, the accuracy of FCTNet with the SGAD framework decreases from 85.6\% (LOTO) to 82.9\% (LOAO) and 80.5\% (LOSpO), with a comparable pattern observed under subject-independent evaluation. Notably, LOSSO, which we introduce as the most stringent protocol, jointly enforces unseen-subject and unseen-speaker conditions, resulting in the lowest accuracy of 76.7\%. These results highlight the increased difficulty of AASD under realistic cross-condition generalization and emphasize the importance of rigorous evaluation protocols for practical neuro-steered hearing systems.

\section{Conclusion}
This study addresses decision instability and evaluation bias in dynamic AASD by introducing a state-guided adaptive decision (SGAD) framework together with structured hierarchical evaluation protocols. By adaptively modulating temporal smoothing according to inferred transition states, SGAD mitigates the trade-off between stability and responsiveness. The consistent performance degradation observed under increasingly stringent protocols further highlights the importance of comprehensive evaluation across audio, speaker, and subject dimensions. Together, these contributions advance the development of reliable neuro-steered hearing systems in dynamic environments.

\section{Acknowledgments}
This work was supported by the National Key Research and Development Program of China (2025YFF0518003, 2023YFF1203502), the National Natural Science Foundation of China (62371217), and the Center for Computational Science and Engineering at Southern University of Science and Technology.

\section{Generative AI Use Disclosure}
All (co-)authors are responsible and accountable for the work and content of the paper, and consent to its submission. Generative AI tools were used for editing and polishing manuscripts, but were not used for producing a significant part of the manuscript.

\bibliographystyle{IEEEtran}
\bibliography{mybib}

\end{document}